\documentclass[11pt]{article}%
\usepackage{amsmath}
\usepackage{geometry}
\usepackage{amsfonts}
\usepackage{amssymb}
\usepackage{todonotes}
\usepackage{graphicx}
\usepackage{authblk}
\providecommand{\U}[1]{\protect\rule{.1in}{.1in}}

\begin{document}

\title{A mathematical model for growth of solid tumors and combination therapy with an application to colorectal cancer}
\author[$\dagger$]{Valeria De Mattei}
\author[$\dagger$]{Franco Flandoli}
\author[$\dagger$]{Marta Leocata}
\author[$\dagger$]{Maria Cristina Polito}
\author[$\ddagger$]{Cristiano Ricci}
\affil[$\dagger$]{Department of Mathematics, University of Pisa}
\affil[$\ddagger$]{Department of Mathematics and Computer Science, University of Florence}
\maketitle

\begin{abstract}
We present a mathematical model, based on ordinary differential equations, for the evolution of solid tumors and their response to treatment. Specifically the effects of a cytotoxic agent and a monoclonal antibody are included as control term in the equations.
The variables considered here are: the number of cancerous cells sensitive to chemotherapy, the number of cancerous cells resistant to chemotherapy, the degree of angiogenesis  and the average intensity of VEGF.  The rules that govern the quantities mentioned above are based on a geometrical argument: we approximate the tumor mass as a sphere and thus derive basic formulae for the normoxic cells and for VEGF production. The monoclonal antibody acts on VEGF and thus has influence to the global degree of angiogenesis. Numerical estimates on some of the parameters are performed in order to match the main landmark in tumor progression and reaction to treatment in the specific case of colorectal cancer. 
\end{abstract}

\section{Introduction}

Mathematical oncology promises to be a powerful tool to study cancer progression. Differential equations models for various quantities, including a spatial structure or not, have been developed with the specific purpose of analyzing the change in tumor size over time. In addition, mathematical models are also helpful to understand the long term dynamics of tumor growth as well as to test biological hypothesis or to evaluate different treatment options. Various technique for quantitative estimates are applied in order to match quantitative values proper of this disease: this is possible after a rigorous calibration of such models using clinical or experimental data. 

Several mathematical models of tumor growth based on ordinary differential equations have been proposed in literature. The core of these models is the presence of two equations for the number of cells and the carrying capacity, that is the level of angiogenesis. The models can reproduce the main features of tumor growth, see for instance \cite{BenzekryHubert}, \cite{Hahnfeld}, where tumor growth is gompertzian, or \cite{Simeoni}, with a modification of gompertzian growth, or  \cite{DonofrioGandolfi}, where the tumor growth is logistic.
Some of these models based on ODE do not omit the spatial structure completely: some of them apply a spherical approximation of the tumor mass ({multicellular tumor spheroids}) and 
prescribe a growth rule based on such approximation, see \cite{Landry} for an example. 

VEGF has already been introduced in previous models: in \cite{Gabhan}, \cite{Jain}, \cite{Stefanini} molecularly detailed models of VEGF have been studied; \cite{Araklein} and \cite{Agur} introduce ODEs models involving the number of cells, angiogenesis and growth factors. 
In literature several models involving cytotoxic and antiangiogenic therapies can be found: see \cite{BenzekryHubert}, where combination between cytotoxic and antiangiogenic therapies is being optimized, \cite{BenzekryBenabdallah}, where the cytotoxic and antiangiogenic drugs are combined with surgery, \cite{Klamka} and \cite{Ledzewicz}, again comparing cytotoxic and antiangiogenic agents,  and \cite{Argyri}, where a model involving antiangiogenic therapy is studied. Moreover, models which involve cytotoxic agents impose a complex argument: birth and evolution of drug resistance cells, widely studied in \cite{ColdmanMurray}\cite{Drake}\cite{Komarova}  \cite{Sadahiro}\cite{Tlsty}.

Inspired by previous works we aim to formulate a mathematical model that is able to match the main landmarks in tumor development in the growth phase as well as the response to specific treatment, with a reasonable level of approximation. 
Summarizing, the general aims of this paper are the following: (1) we want to develop a model to reproduce qualitative aspects of the growth, specifically the transition between the exponential growth in the beginning and a slower regimen when the number of cells increase, trying to partially restore the exponential growth after the angiogenesis takes place. (2) We would like to specialize our model to a concrete case (colorectal cancer) and calibrate the parameters in order to match quantitative values proper of this disease. (3) Referring to the previous point, we want to embed in our model a combination therapy in order to compare the results obtained with those coming from clinical trials. Specifically, we have in mind to include the action of a cytotoxic agent, directly on the number of cells, as well as the action of an antiangiogenic agent on the VEGF variable. We also plan to introduce  drug-resistant cells into the system to take into account the deterioration of chemotherapy efficacy.

The paper is organized as follow. The mathematical model is presented in section \ref{section model 1}. Section \ref{section parameters} is entirely devoted to the choices for the parameters, whose calibration was performed in the specific case of colorectal cancer, while in section \ref{sec::numericalsimulations} we present the numerical results concerning size and TTP (time to progression) in a specific treatment regimen. Finally in section 5, a brief conclusion is given.

\section{The mathematical model\label{section model 1}}

The general rationale of our model construction has been to use a very small
number of variables, with a clear biophysical meaning, simple but still
preserving general main mechanisms of cancer growth.
We neglect details of the spatial structure and thus do not consider
the position of cancer cells and the space-dependence of VEGF concentration and vascularization.
We consider the following variables:

\begin{itemize}
\item $N_{t}^{sens}$ = number of drug-sensitive cells;

\item $N_{t}^{res}$ = number of drug-resistant cells;

\item $V_{t}$ = intensity of VEGF field (space average);

\item $A_{t}$ = level of vascularization due to
angiogenesis (space average).
\end{itemize}
For convenience we introduce also the total number of cells:
\begin{itemize}
\item $N_t = N_t^{sens}+N_t^{norm}$.
\end{itemize}
In order to include the effect of a cytotoxic agent (5-FU) we divide the population of cancerous cells into two subpopulations: here $N_t^{sens}$ represents the number of cells sensitive to chemotherapy and $N_t^{res}$ the number of drug-resistant cells. Although the complexity of drug-resistance phenomenon, we assumed for simplicity the resistance as a binary property.
To describe the dynamic of the two populations we follow the classical approach of distinguishing the proliferative behavior according to oxygen availability. Thus we divide each population of cells into two subclasses: normoxic (hence potentially proliferating) and hypoxic cells, denoted by $N_t^{i,norm}$ and $N_t^{i,hypo}$ respectively.

Let us now describe the differential equations
satisfied by the previous quantities:
\begin{equation}
\label{eq::N^{sens}}
\hspace{-2.8cm}\frac{d}{dt}N_{t}^{sens} =(1-p)\cdot (1-u_t^{FU})\cdot\lambda N_t^{sens,norm}-\mu N_{t}^{sens}
\end{equation}
\begin{equation}
\label{eq::N^{res}}
\hspace{-1.9cm}\frac{d}{dt}N_{t}^{res} =p\cdot(1-u_t^{FU}) \cdot\lambda N_t^{sens, norm}+\lambda N_t^{res,norm}-\mu N_{t}^{res}
\end{equation}
\begin{equation}
\label{eq::V_{t}}
\frac{d}{dt}V_{t}=\left(  C_{hypo\rightarrow V}\left(  \frac{N_{t}^{hypo}}{N_{t}}\right)  ^{2/3}-C_{A,V}A_{t} - C_{beva\rightarrow V}u_t^{beva}\right)  V_{t}\left(1-V_{t}\right)
\end{equation}
\begin{equation}
\label{eq::A_{t}}
\hspace{-1.5cm}\frac{d}{dt}A_{t}=1_{V_{t}>V_0}C_{V\rightarrow A}\left(  V_{t}-V_0\right)\left(  A_{t}+A_0\right)  \left(  1-A_{t}\right)  -2A_{t}^{k_A}1_{V_{t}\leq V_0}
\end{equation}
where

\begin{equation}
\label{eq::N^{hypo}}
N_{t}^{i,hypo}=\left(  1-A_{t}\right)  N_{t}^{i}\left(  1-\frac{1}{1+\left(  N_{t}^{i}\right)^{1/3}/2\eta_{i}}\right)^{3}
\end{equation}
and $u_t^{FU}$ and $u_t^{beva}$ describes the tissue concentration of the cytotoxic agent and of the monoclonal antibody respectively (more details in section \ref{subsec::kabbinavar}).
\subsection{On the geometrical configuration of the hypoxic and normoxic cells\label{subsection::geometrical}}
Let us now introduce the spherical approximation that we have used to describe the variable fraction of the normoxic (resp. hypoxic) cells over the total number of cells. We assume that there
are two classes of normoxic cells: those near the boundary of the tumor mass, and those not near the boundary but reached by the new vascular network produced
by the angiogenic cascade. In order to have a simple formula for the boundary of the tumor, we assume it has a spherical shape - this is true only approximately, for various reasons, including the complex geometry of real tissues and the rugosity of the surface, but we cannot include these factors unless we go back to space-dependent models. Let $\delta$ be the thickness of the proliferating boundary and let $R$ and $r$ be the radius of the tumor mass and of a single tumor cell respectively. By a simple calculation we have
\[
N^{norm}_{boundary} = N\left( 1-\left( 1-\frac{\delta}{R}\right)^{3}\right)
\]
Let us consider two extreme regimes regarding tumor size.  When the tumor is very small, $\delta
=R$. When the tumor is very large, we assume $\delta$ stabilizes to a certain
number $\eta$ of cell diameters, hence $\delta=2\eta r_{cell}$. Therefore%
\[
\frac{\delta}{R}=\left\{
\begin{array}
[c]{cc}
1 & \text{when }N\text{ is small}\\
2\eta/N^{1/3} & \text{when }N\text{ is large}
\end{array}
\right.  .
\]
A simple function which interpolates these two extremes is
\[
\frac{\delta}{R}=f\left(  N,\eta\right)  =\frac{1}{1+N^{1/3}/2\eta}.
\]
Thus, we arrive to the formula:
\[
N_{boundary}^{norm}=N\left(  1-\left(  1-\frac{1}{1+N^{1/3}/2\eta}\right)
^{3}\right).
\]
The contribution coming from those cells, in the complementary of the boundary
layer, which are sufficiently angiogenized to be considered as normoxic, is
given by
\[
N_{angio}^{norm} = A\cdot N \left(  1-\frac{1}{1+\left(  N\right)
^{1/3}/2\eta}\right)  ^{3} \label{contrib angio},
\]
where $A_{t}$ is the average level of vascularization defined above. The hypoxic cells are simply obtained by difference. 
\subsection{Growth dynamics of sensitive and resistant cells} 

The reason of deterioration of therapy we take into account is the presence of drug-resistant mutations. We shall rewrite equation \eqref{eq::N^{sens}} and \eqref{eq::N^{res}} using vector notation: 
\begin{equation}
\label{eq::mutate}
\frac{d}{dt}N_{t} = \lambda U P^{t} N_{t}^{norm} - \mu N_{t}
\end{equation}
where 
\[
N_{t} = \binom{N_{t}^{sens}}{N_{t}^{res}},\quad N_{t}^{norm} = \binom{N_{t}^{sens,norm}}{N_{t}^{res,norm}}, \quad P=\begin{bmatrix}
1-p & p \\
0 & 1
\end{bmatrix}, \quad U=\begin{bmatrix}
1-u_{t}^{FU} &0 \\
0 & 1
\end{bmatrix}.
\]
We simplify and prescribe that each cancer cell, during a duplication, has a probability 
$p$ of developing a drug-resistant mutation. The mechanism of mutation is like a change of species. Following this general rationale we summarized this phenomenon by the transition matrix $P$ of a Markov chain.
A precise rule of growth of the two subpopulations $N_{t}^{sens}%
,N_{t}^{res}$ would be quite intricate and depend on the relative geometry.
Indeed, the approximation of a spherical shape cannot work simultaneously for
both, since they are interlaced. Let us distinguish before and after treatment.
Before treatment, $N_{t}^{sens}>>N_{t}^{res}$, hence the sphere approximation
for drug-sensitive is reasonable. But
drug-resistant cells are distributed, inside the tumor mass, in a quite
complex way. Somewhere we expect to see kernels of drug-resistant cells due to
a first mutated cell together with its descendants;\ however, this kernel
cannot proliferate indefinitely as the global tumor mass because it is partially
or completely absorbed into the deeper, hypoion. In order to ``close'' our equations, we conjecture that the proliferating boundary of the drug-resistant cells is thinner than the case of sensitive cells. For this reason, and for the sake of simplicity,  we use the same spherical approximation even for the resistant cells but we take $\eta_{res}<\eta_{sens}$.
After treatment, the approximation as two separate spheres becomes more
reasonable, since mutated kernels of cells may remain isolated by the loss of
surrounding non-mutated cells and, again to simplify, we assume that the
drug-resistant mass is dominated by the largest kernel of mutated cells, so
the approximation as a single sphere is not too distant from reality. In this
case we thus take $\eta_{res}=\eta_{sens}$.

We point out that equation \eqref{eq::mutate} contains a small inaccuracy: during treatment sensitive cells are affected by chemotherapy while resistant cells are not. In this phase the control term $1-u_{t}^{FU}$ is negative and causes a decrease in the number of sensitive cells: this decrease is slightly altered due to the fact that $N_{t}^{sens,norm}$ is scaled by the term $(1-p)$ coming from the matrix $P$. In practise this inaccuracy will be numerically negligible given that average values of $p$ will range from $10^{-6}$  to $10^{-4}$ and thus $(1-p) \sim 1$.

\subsection{The $2/3$ formula for VEGF and the angiogenesis (de)generation}

The VEGF field is produced by hypoxic cells and absorbed by endothelial cells.
The factor $V_{t}\left(  1-V_{t}\right)  $ in equation \eqref{eq::V_{t}} has the role to restrict $V_{t}$ in
$\left(  0,1\right)  $, where the choice of $1$ as maximal value of $V_{t}$ is
conventional. The term $-C_{A,V}A_{t}$ is the natural one to describe
absorption by endothelial cells - moreover, it will play a very marginal role
in the sequel. 

The $2/3$  rule for VEGF production by
hypoxic cells is justified by the argument that follows.
Consider the sphere of hypoxic cells as it were a uniformly ``charged'' sphere
of radius $R-\delta$:\ it induces the electric field
\[
E(r)=\frac{C\left(  R-\delta\right)  ^{3}}{r^{2}},
\]
for a suitable constant $C$. The potential on the boundary of the sphere is%

\[
V(R)=C\left(  R-\delta\right)  ^{2}.
\]
This is the work done by the electric field to move a point charge from
infinity to the boundary of the sphere. In analogy, we think to VEGF as the
work needed to move blood vessels from an infinite distance to the sphere of
hypoxic cells. Thus VEGF would have the form

\[
V_{t}=C\left(  R_{t}-\delta_{t}\right)  ^{2}=\widetilde{C}(N_{t}^{hypo})^{2/3}%
\]
with a new constant $\widetilde{C}$.

The equation \eqref{eq::A_{t}} describes the growth of vasculature by the term
\[
1_{V_{t}>V_0}C_{V\rightarrow A}\left(  V_{t}-V_0\right)\left(  A_{t}+A_0\right)  \left(  1-A_{t}\right),
\]
while the second term 
\[
-2A_{t}^{k_A}1_{V_{t}\leq V_0}
\]
concerns the regression due to monoclonal antibody. The bio-mechanical rule of regression is different from the one of accretion. Indeed, when regression occurs, the more recent
capillaries are still poor in structure - e.g. not so covered by pericytes -
and are easily destroyed, in a short time (of the order of a week) compared
with the long times of the overall process (of the order of many years). On
the contrary, the less recent angiogenic vasculature has already reached a
certain level of stability and thus it regresses much slowly. We have devised,
with some degree of approximation and of ad-hoc numerical trials, the
above law for the regression.

The term $A_{0}$ has been added because for values of $A_{t}$ very close to
zero - even equal to zero - the growth $\frac{d}{dt}A_{t}$ is not
infinitesimal, but finite non zero, due to the presence of the pre-existing
vasculature.

The term $V_{0}$ corresponds to the fact that angiogenesis requires some
degree of concentration of VEGF. Below such threshold no angiogenesis
occurs;\ or more precisely, a regression takes place.

\section{Parameters\label{section parameters}}

\begin{table}[!ht]
\centering
\begin{tabular}{p{0.5cm}p{7cm}p{1cm}p{3.5cm}}
\multicolumn{1}{l}{}                    &                                &&                                 \\ \hline
\multicolumn{1}{c}{\textbf{Parameters}} & \multicolumn{1}{l}{\textbf{Meaning}}              &\multicolumn{1}{l}{\textbf{Value}}   &    \multicolumn{1}{l}{\textbf{Source}}              \\ \hline
$\lambda$                                & growth rate due to cell proliferation         &0.05&\cite{Bolin}\cite{Friberg}\cite{Sadahiro}\cite{Umetani}   \\ 
$\mu$                                   & decay rate due to cell loss                            &0.002&\cite{Bolin}\cite{Friberg}\cite{Sadahiro}\cite{Umetani}       \\ 
$\eta_{sens}$                           & thickness of proliferating boundary of drug-sensitive cells  &15&\cite{Weinberg}  \\ 
$\eta_{res}$                            & thickness of proliferating boundary of drug-resistant cells  &$\frac{\eta_{sens}}{1.2}$&  estimated/simulations \\ 
$C_{hypo\rightarrow V}$                 & VEGF production rate from hypoxic cells              &0.08&      estimated/simulations        \\ 
$C_{A,V}$                               & absorption rate of VEGF from vasculature              &0.01&    estimated/simulations        \\ 
$C_{V\rightarrow A}$                    & reaction rate of angiogenesis to VEGF              &0.006&    estimated/simulations           \\ 
$A_0$                              &  preexistence level of vascularization      &0.2&         estimated/simulations       \\ 
$k_A$                              & angiogenesis regression rate    &10&      estimated/simulations          \\ 
$V_0$                              & VEGF threshold for angiogenesis progression &0.2&     estimated/simulations           \\ 
$p$                                     & probability of drug resistant mutation per cell per duplication &$10^{-5}$&\cite{ColdmanMurray}\cite{Drake}\cite{Komarova}  \cite{Sadahiro}\cite{Tlsty}\\
$N_{start}$                             & number of cells when therapy starts                        &$10^{9}$& \cite{Machida}\cite{Zhao}     \\ 
$N_{FU}$                                & number of days of action of 5-FU (1 to 7)                &6&  \cite{Peters}      \\ 
$C_{FU}$                                & intensity of 5-FU action                                 &20& estimated/simulations  \\ 
$N_{beva}$                              & number of days of action of BV5 (1 to 14)         &12&  \cite{Gaudreault}\cite{Motl}             \\ 
$C_{beva\rightarrow V}$                 & inhibition rate of bevacizumab on VEGF        &1& estimated/simulations    \\ \hline
\end{tabular}
\caption{{\protect\small Parameters of the model. Parameters with no source have been estimated indirectly, refer to section \ref{subsec::kabbinavar}.}}
\label{tab::parammathmodel}
\end{table}
This section specializes our model to the case of colorectal cancer.  As one can appreciate from table \ref{tab::parammathmodel}, some values of parameters are available in literature while others, harder to measure, are estimated experimentally. The first class will be discussed here, while the others are argued in section \ref{subsec::kabbinavar}.

\begin{figure}[ptb]
\centering
\includegraphics[height=3.1168in,width=4.152in]
{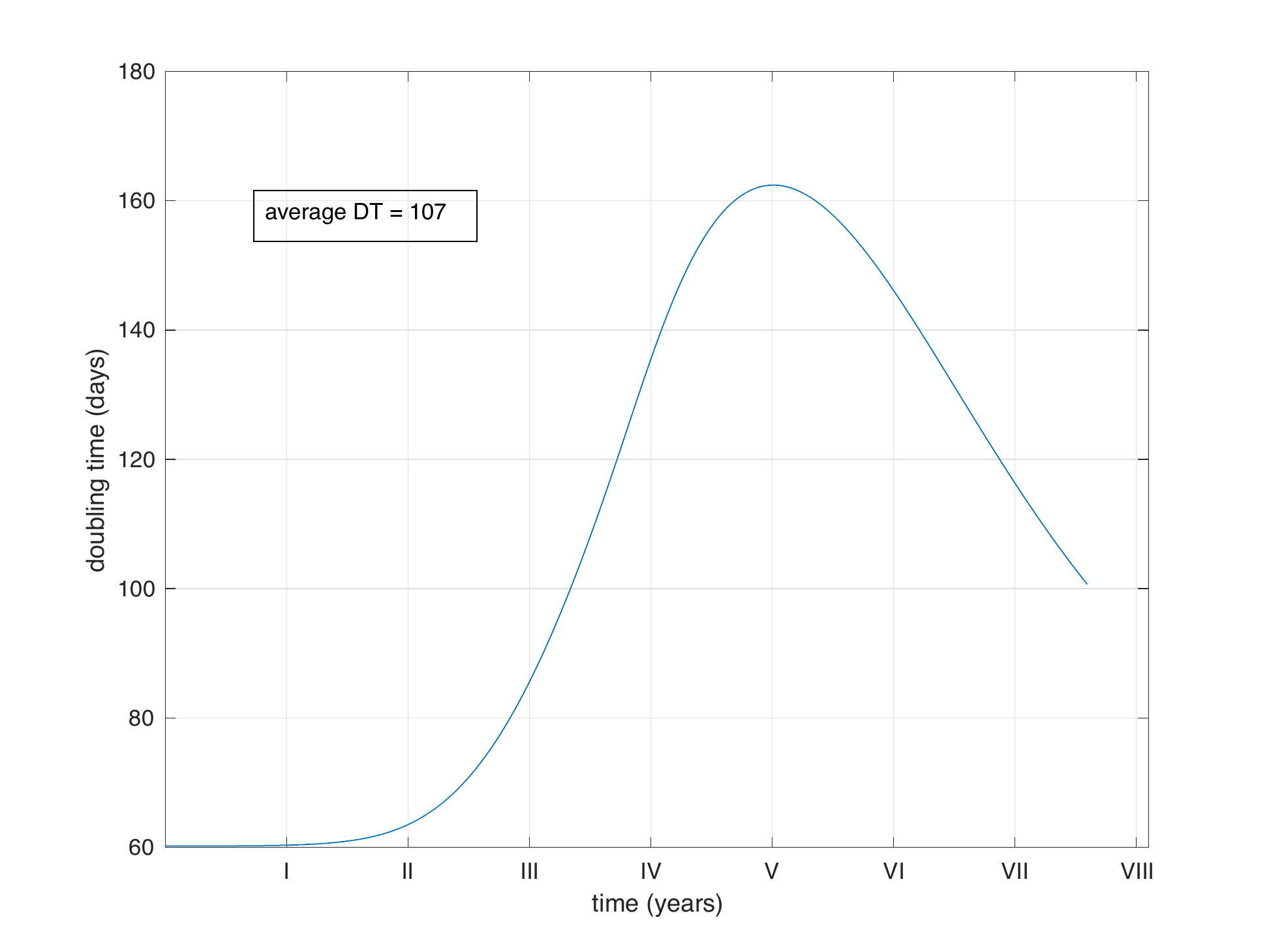}
\caption{{\protect\small Istantaneous doubling time (DT) as it varies over the
whole period. The average is 107, coherent with the literature.}}
\label{fig DT}
\end{figure}

We based our approximation of $(\lambda, \mu)$ on the estimate of doubling time of colorectal cancer. From the literature (see for instance \cite{Friberg} \cite{Bolin}, \cite{Sadahiro}, \cite{Umetani}) DT of the order of 60-180 days seems possible (also less and more). Some authors (see \cite{Steel}, \cite{Bianco}) claim that periods of
the order of 8 years could be reasonable estimates of the time needed to reach
$10^{9}$ cells and this result is not incompatible with the previous estimates
of DT, up to some degree of approximation. We have chosen to impose roughly 8
years to our model to reach $10^{9}$ cells and we have computed an average DT
of the order of 100 days, where by average we mean that we take into account
the changes of speed of growth that our model has, we compute at every time
the instantaneous doubling time and we average it (fig. \ref{fig DT}). Obviously there are several choices of parameters which produce these results. We choose as a standard the values in table \ref{tab::parammathmodel}.

The number of cells when therapy starts is obtained by translating available statistics about the volume which around tumors become detectable ($\sim 1 cm^3$), see for instance \cite{Machida} \cite{Zhao}. 

\begin{figure}[]
\centering
\includegraphics[height=3.1168in,width=4.152in]
{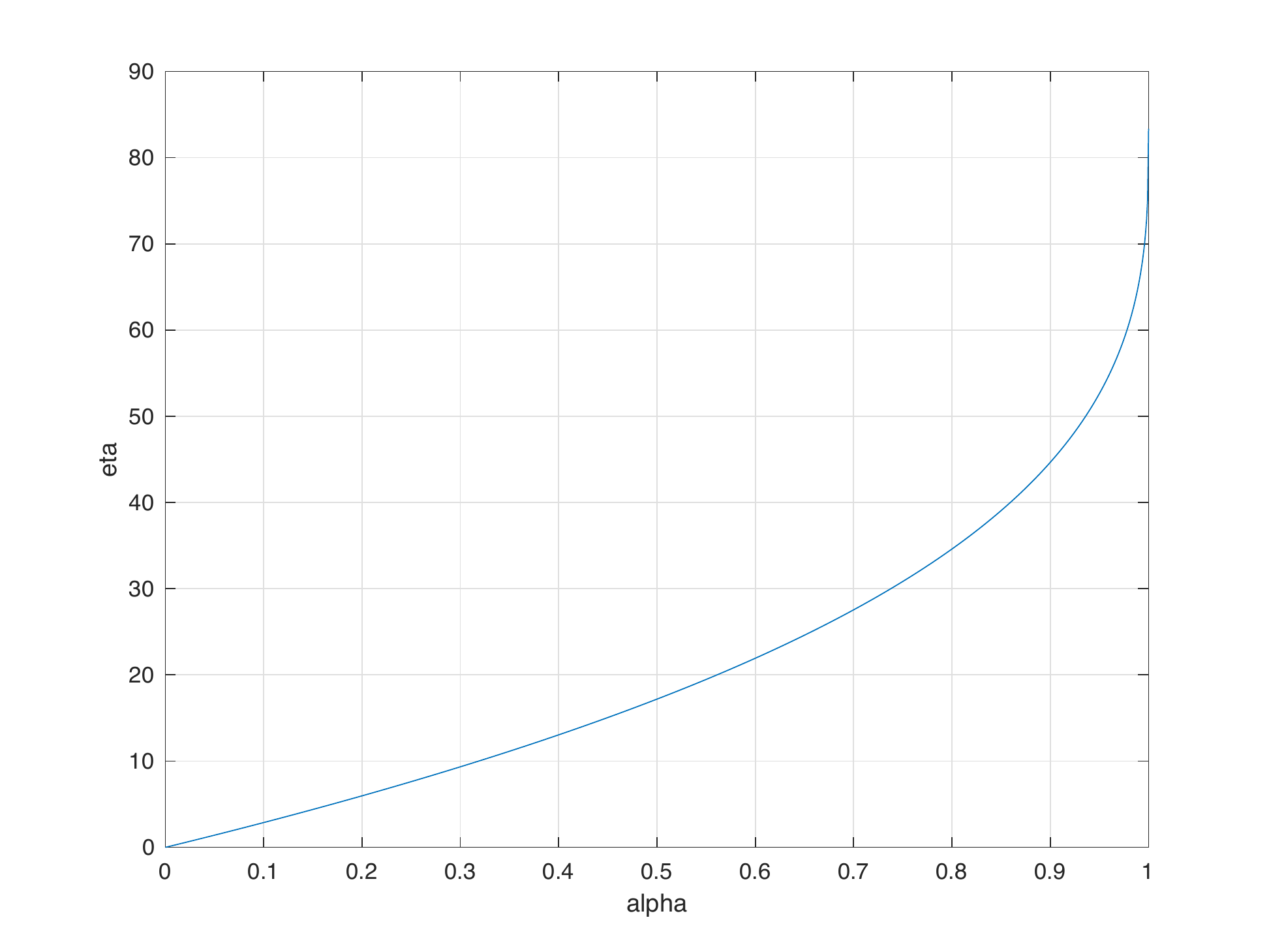}
\caption{{\protect\small Size of }$\eta${\protect\small  \ as a function of
}$\alpha${\protect\small , see text.}}
\label{fig eta}
\end{figure}

For what concerns the thickness of proliferating boundary $\eta_{sens}$, we obtain the range $5-30$ through two different arguments. The first one is given by experimental estimates about the distance from a capillary at which cells start to be hypoxic or even necrotic; see a collection of results in \cite{Weinberg}, figure 13.27 and references therein. These estimates range between 60 and 110 $\mu m$, namely less than 10 cells. However, the external proliferating layer is exposed to a much richer amount
of oxygen and nutrients than a packed tumor around a capillary. Thus, the
previous estimates should be corrected in the increasing direction.
Another argument comes from the observation that tumors with radius of size
around $1mm$ are usually avascular, hence for them the proportion $\alpha$ of
proliferating boundary with respect to the total is not small. Let us see that
values of $\alpha$ of the order 0.25, 0.50, 0.74 give rise to estimates of
$\eta_{sens}$ again of the order of 10. We assume as above that the cell
radius is $6\mu m=6\cdot10^{-3}mm$. Similarly to the computation of the section \ref{subsection::geometrical} we assume to have a spherical tumor of radius $R$ and
an external proliferating layer of thickness $\delta$;\ hence, in $mm$,
$\delta=2\cdot6\cdot10^{-3}\eta_{sens}$. We want to estimate $\delta$, and
therefore $\eta_{sens}=10^{3}\delta/12$. The number $\alpha$ is the ratio
between the two volumes, hence
\[
\alpha=\frac{R^{3}-\left(  R-\delta\right)  ^{3}}{R^{3}}=1-\left(
1-\delta\right)  ^{3}%
\]
where we have used $R=1$. Hence%
\[
\eta_{sens}=\frac{10^{3}\left(  1-\sqrt[3]{1-\alpha}\right)  }{12}.
\]
We see from figure \ref{fig eta} that reasonable values are around 10-30;
mixing with \cite{Weinberg} we choose $\eta=15$.

The value of $p$ is a very delicate issue, see \cite{Drake}: it derives from the assumption that the mechanism of mutation is a binary random variable. 
Our reference value has been $10^{-5}$ and the range under randomization has been between
$10^{-4}$ and $10^{-6}$: we obtained this values by numerical simulations comparing our results with the clinical about TTP  found in \cite{Kabbi2005}, more details in section \ref{subsec::kabbinavar}. In support of these values let us quote for instance
\cite{Tlsty}, \cite{Goldie} (page 3650), \cite{Komarova} (section 5.3),
\cite{ColdmanMurray}. The long time between cell divisions of colorectal cancer cells may
be a factor of increase of $p$, see \cite{Gao}. 

The parameters $N_{FU}, C_{FU}, N_{beva},C_{beva\rightarrow V}$ are included in the control functions $u_t^{FU}, u_t^{beva}$. Concerning 5-FU, its half-life in plasma is very short, of the order of 10-20
minutes, but in tissues occupied by cancer cells appreciable levels are
measured for days, see \cite{Peters} (in particular figure 1). Concerning BV5, it seems it remains active for a long period, see for instance \cite{Gaudreault}\cite{Motl}, thus we take $N_{beva}=12$. 
The study of parameters, whose values are not known in literature, represents one of the most difficult point of our work: the idea was to search for the range of admissible values through numerical simulations. Fixing some constraints we have studied the reasonable value of parameters. As largely explained in the following section, we have in mind two kind of constraints: the first class consists of universal characteristics of tumor evolution while the second consists of specifics of colorectal cancer.

\begin{figure}[ptb]
\centering
\includegraphics[width=\textwidth, height=7.9cm]{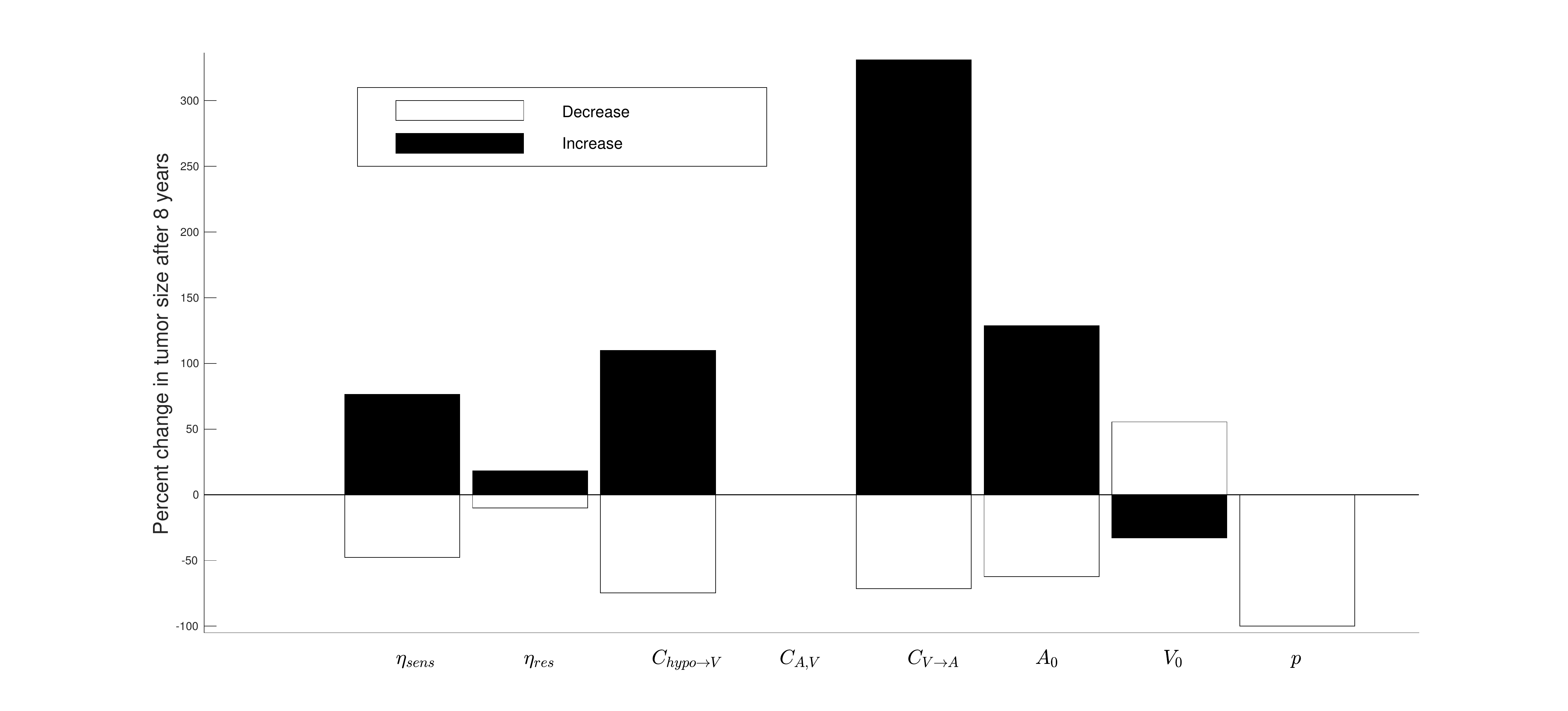}
\caption{{\protect\small Sensitivity analysis: parameters changed vs percentage change in tumor size after 8 years of growth. Parameters are changed of 50\% from table \ref{tab::parammathmodel}, except $p$, which is altered from $10^{-6}$ to $10^{-4}$.}}
\label{sens:: size}
\end{figure}

\begin{figure}[ptb]
\centering
\includegraphics[width=\textwidth, height=7.9cm]{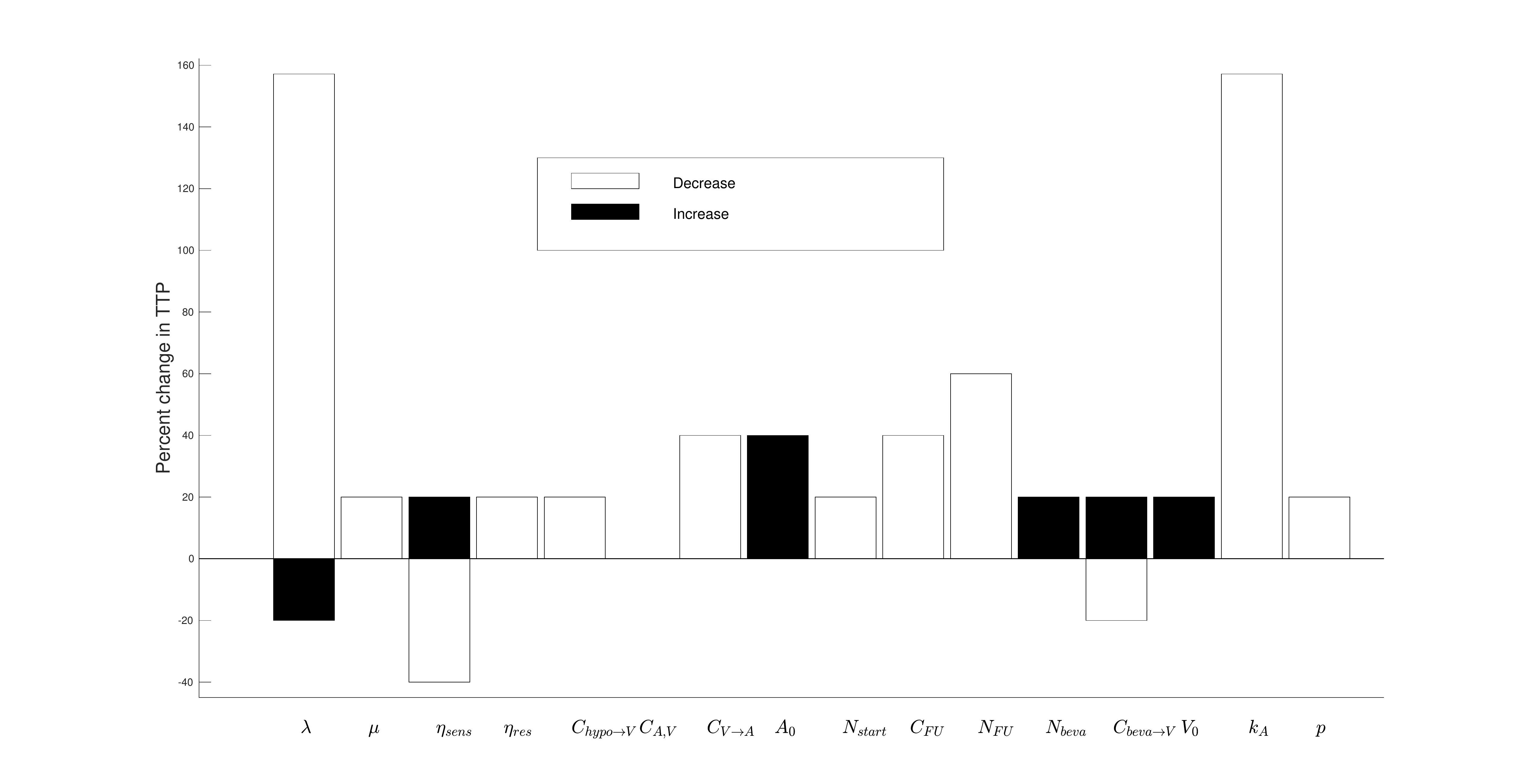}
\caption{{\protect\small Sensitivity analysis: parameters changed vs percentage change in TTP after treatment (see section \ref{subsec::kabbinavar}). Parameters are changed of 50\% from table \ref{tab::parammathmodel}, except $p$ and $N_{start}$, which are altered from $10^{-6}$ to $10^{-4}$ and from $10^{8}$ to $10^{10}$ respectively.}}
\label{sens::TTP}
\end{figure}

\section{Numerical Simulations}
\label{sec::numericalsimulations}
In this section we propose two different approaches to investigate and validate the model. In the first section we show the biological events caught by our model, while in the second, focused on therapy, we try to compare our result with clinical data proposed in \cite{Kabbi2005}. 
\begin{figure}[h]
\centering
\includegraphics[height=3.1159in,width=5.5694in]
{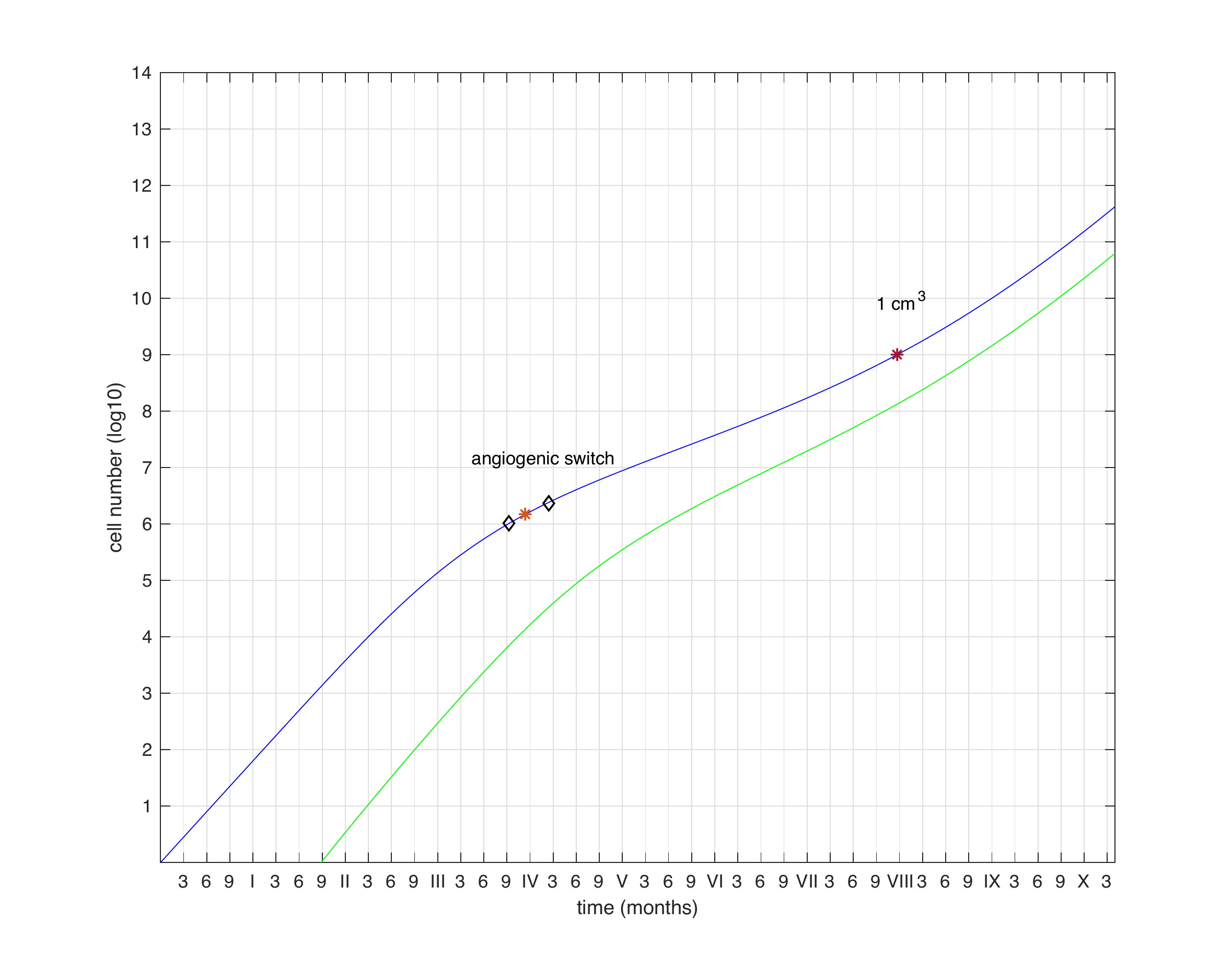}
\caption{{\protect\small Simulation without therapy. Diamonds
denote the range} ${\protect\small 10}^{6}{\protect\small -2\cdot10}^{6}$ where angiogenesis is expected to start and the first red
star is that initial time. The second red star is when $10^{9}$ cells are reached which should be close to 8 years. Blue line denotes total number of cancer cells, green line those
having a drug resistant mutation. Concerning parameters, their value for this
simulation is given in  table \ref{tab::parammathmodel} of sections \ref{section parameters}.}
\label{fig::main}
\end{figure}
\subsection{General shape of growth}\label{subsec::shape}
By observing figure \ref{fig::main} we can appreciate the general shape of the growth curve over a large time interval (roughly ten years). When the tumor is small we have $N^{norm}_{boundary}\sim N$, hence in the first segment of the curve the growth is purely exponential. When the tumor is larger, namely when $N^{1/3}/\eta \gg 1$, we have
\[
N_{layer}^{norm}\sim N\left(  1-\left(  1-\frac{2\eta}{N^{1/3}}\right)
^{3}\right)  \sim3N\frac{2\eta}{N^{1/3}}=6\eta N^{2/3}.
\]
The $2/3$ law often appears in the literature, see for instance \cite{TalkingtonDurrett} and the references therein. Later on, thanks to the angiogenic contribution, we restore the exponential growth rate: in fact when $N$ is large 
\[
N\left(1-\frac{1}{1+N^{1/3}/\eta}\right)^{3} \sim N.
\]

Comparing the solution of the system \eqref{eq::N^{sens}}-\eqref{eq::A_{t}} with a classical Gompertz model, characterized by only two phases, it is possible to distinguish a third phase where the exponential rate is restored by angiogenesis.

Keeping in mind the focus on colorectal cancer, we direct our attention to the numerical values proper of this disease: 
\begin{itemize}
\item roughly 8 years to reach $10^{9}$ cells $\sim 1cm^{3}$;
\item angiogenesis starting between $10^{6}$ and $2\cdot 10^{6}$.
\end{itemize}
The first landmark of $1 cm^{3}$ has already been argued in section \ref{section parameters}. Concerning the starting of the angiogenic cascade, $1mm^{3}$ is a typical order of magnitude of the largest avascular tumors. Around $1-2$ $mm^{3}$ the angiogenic cascade starts, see \cite{Marme}. This translates to the range $10^{6}-2\cdot10^{6}$. With these values in mind we have chosen the parameters $C_{hypo\rightarrow V}, C_{A,V}, C_{V\rightarrow A}, A_{0}$ in order to reproduce these specific events. Obviously there are several choices of parameters which produce these results (see figure \ref{fig::main}). We choose as standard the values in table \ref{tab::parammathmodel}.
To deepen how the just mentioned parameters influence the model, we propose the analysis of sensitivity in figure \ref{sens:: size}. From this analysis one can appreciate the fact that, increasing (resp.  decreasing) of $50\%$ parameters which calibrate rate of angiogenesis development  ($C_{hypo\rightarrow V},C_{V\rightarrow A}$), implies a relevant increase (resp. decrease) of tumor size. The same linear effect arises from the parameters that influence the rate of proliferation for sensitive and resistant cells ($\eta_{sens},\eta_{res}$). Given the strict correlation between $(\lambda, \mu)$ they have not been tested in this phase. In conclusion, the shape of growth curve seems to be very stable with respect to parameters: their alteration entails the maintenance of the three phases, above mentioned. 
\begin{figure}[p]
\centering
\includegraphics[height=3.4168in,width=5.449in]
{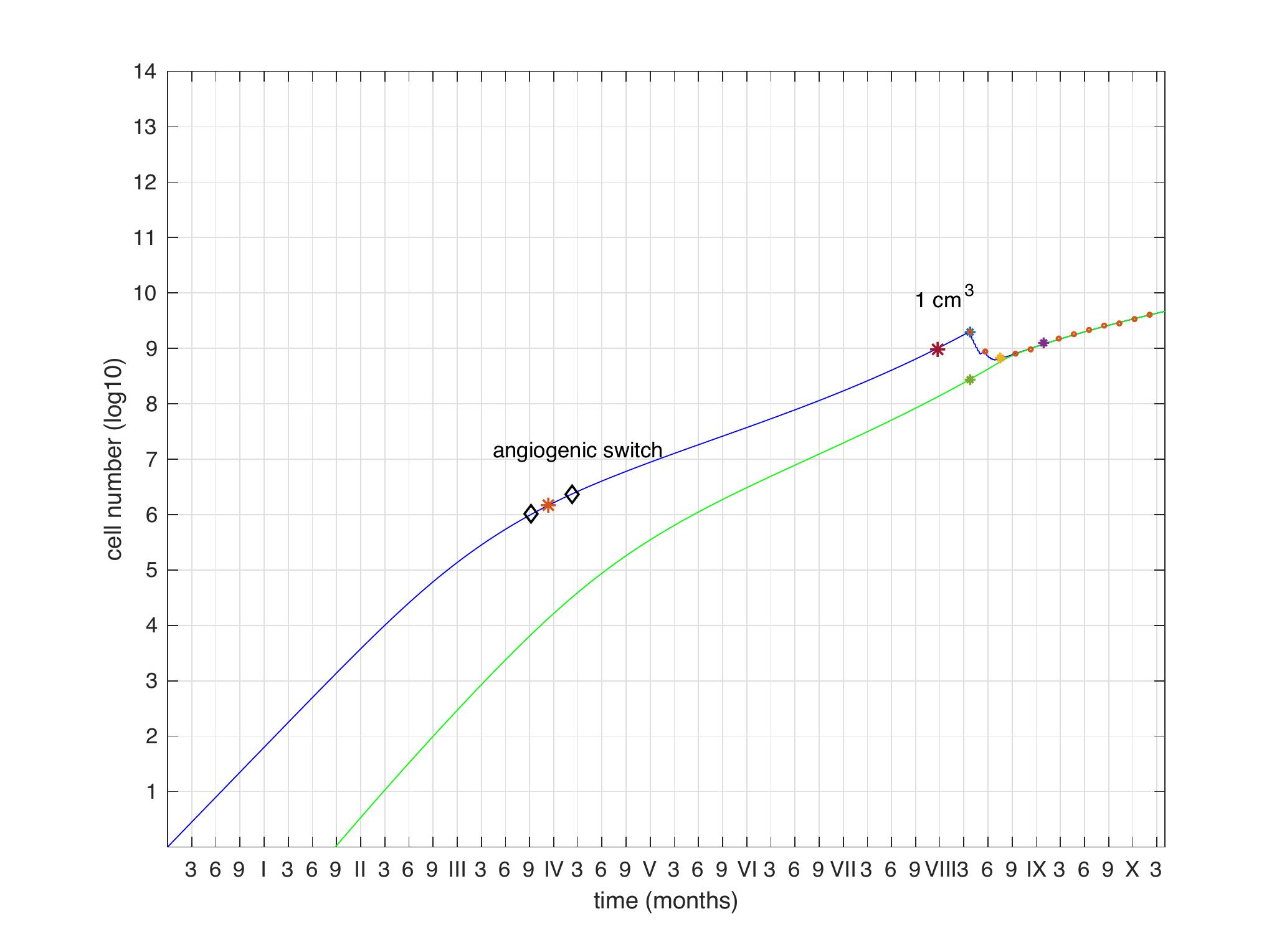}

\end{figure}
\begin{figure}[pb] 
\centering
\includegraphics[height=3.4168in,width=5.4649in]
{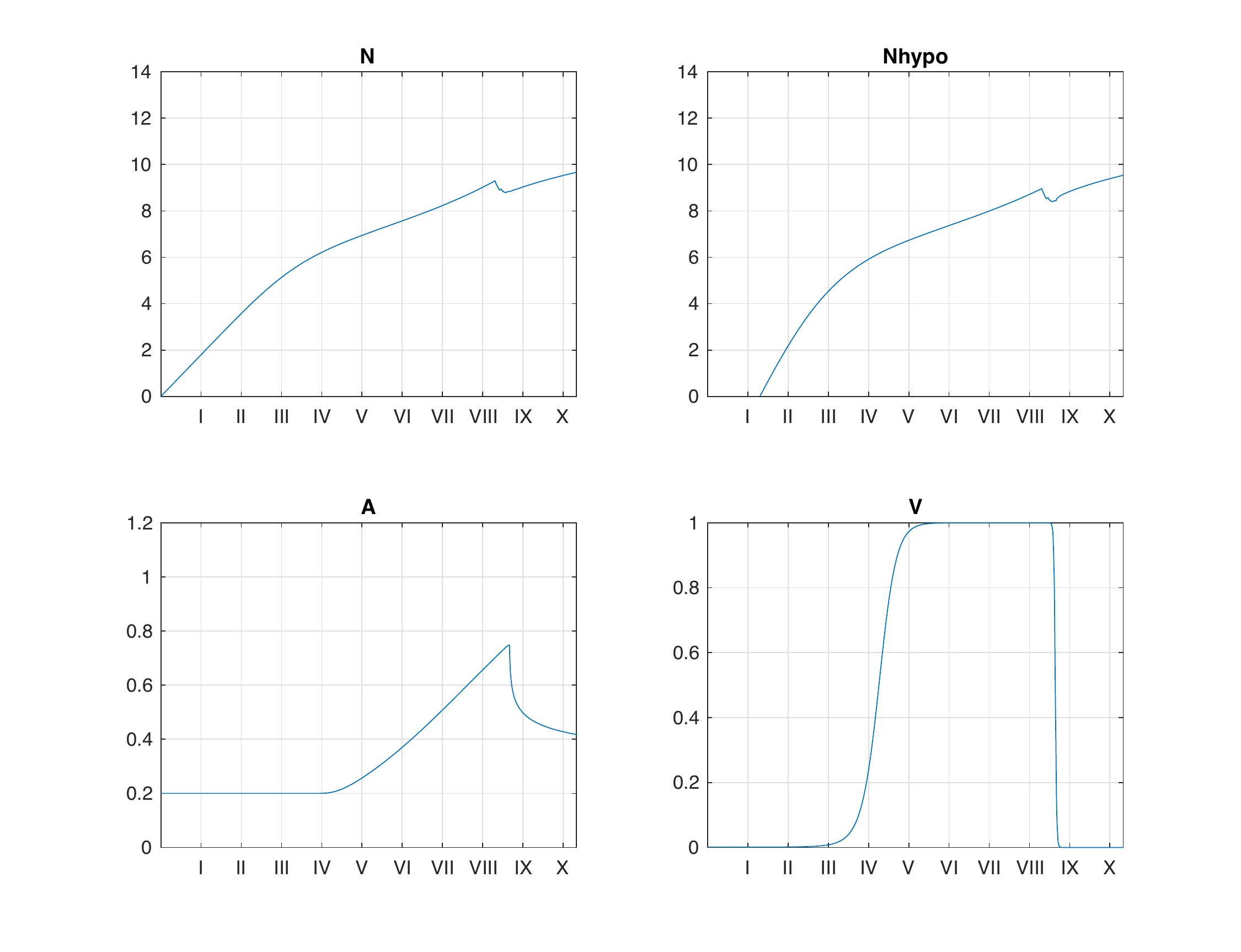}
\caption{{\protect\small Simulation with therapy, 5-FU plus BV5. Circles denote times of assessment of tumor size. Small subgraphs show the behavior of total number of cells, hypoxic cells, degree of angiogenesis and density of VEGF.}}
\label{graf::tot}
\end{figure}

\subsection{Connections with clinical data}
\label{subsec::kabbinavar}
In this section, we follow a second type of validation mechanism for the model: we adapt the model to a specific case and we try to compare the results obtained through simulations with those found in literature.
In clinical trials, when tumor progression takes place, the first line therapy is interrupted and patients could receive any second line treatment, which however may vary and be thus less suitable for a mathematical modeling. Hence, we limit ourselves to a model of the 5-FU or 5-FU + BV5 regimen during the first-line therapy and we measure TTP. Our model is not able to capture the spatial structure of the tumor, so we limit our measurement of TTP on a single lesion of the colorectal cancer. According to \cite{Kabbi2005} and RECIST criteria, we start from time $t_{0}$ when therapy begins and register the number of cells every 8 weeks, namely at the end of each cycle; hence, using the spherical approximation, we convert number of cells into lesion diameter and we compute the first measurement with an increase of at least 20 percent from the minimum registered.
Following \cite{Kabbi2005}, we investigate the case when patients are
initially treated by 12 cycles of 5-FU (plus folinic acid), each cycle being
of 8 weeks with\ 5-FU administrated at the beginning of each one of the first
6 weeks of a cycle. We compare this regimen with the one based on 5-FU (and
folinic acid) plus bevacizumab. This monoclonal antibody is given once every
two weeks - also the last two weeks of the cycle. The particular regimen
bevacizumab 5 mg/kg every 2 weeks will be denoted by BV5 and we refer only to
it when experimental data are quoted. 
We remind that $u_{t}^{FU}$ and $u_{t}^{beva}$ denote the control functions 
corresponding to 5-FU and bevacizumab respectively, functions which are equal
to zero when 5-FU and BV5 are not present in the tissue. To be precise, they
describe the concentration of 5-FU and BV5 in the tumor tissue. Two possible
misunderstandings should be avoided:\ i) the controls $u_{t}^{FU}$ and
$u_{t}^{beva}$ are not understood as the actions of drug administration, which
are short in time like in the case of 5-FU in bolus;\ ii) they correspond to
drug concentration in the tumor tissue, not in plasma. We prescribe that the first administration of 5-FU (+BV5) takes place when the total number of cells reaches the value of $N_{start}$.
Concerning 5-FU, we keep into account two facts:\ i) cell kill by chemotherapy
is faster than proliferation; ii) the concentration of 5-FU in the tissue
decays exponentially in time. Using the following expression:
\[
u_{t}^{FU}=\left(  1+C_{FU}\right)  \exp\left(  -\frac{\log\left(
1+C_{FU}\right)  }{N_{FU}}t\right)  .
\]
we obtain that at the beginning of the week, when 5-FU is given in bolus, the rate is equal
to%
\[
\lambda\left(  1-u_{0}^{FU}\right)  =\lambda C_{FU}.
\]
Hence the constant $C_{FU}$ acts as a multiplier of intensity of cell kill
with respect to cell proliferation. A typical value we use is $C_{FU}=20$. The
expression inside the exponential has the following motivation: at time
$t=N_{FU}$, we have $u_{t}^{FU}=1$, hence the rate at that time is
$\lambda\left(  1-u_{N_{FU}}^{FU}\right)  =0$. In other words, prescribing the
number $N_{FU}$ of days of action of 5-FU, we prescribe that cell loss occurs
until time $N_{FU}$, in an exponentially decreasing manner;\ and afterwords
proliferation restarts but not immediately with full rate $\lambda$, just with
rate $\lambda\left(  1-u_{t}^{FU}\right)  $ which is asymptotic to $\lambda$
for large times.

Concerning bevacizumab, it is given once every two weeks in each
cycle. In the periods of no treatment, $u_{t}^{beva}=0$. Under treatment, we
impose that its half-life is $N_{BV}$, hence the form is
\[
u_{t}^{beva}=\exp\left(  -\frac{\log2}{N_{BV}}t\right)  .
\]
In figure \ref{graf::tot} it is shown the simulation of the therapy, including bevacizumab. Two of nine parameters whose values are not available in literature are included in the control functions, namely $C_{FU}$, $C_{beva\rightarrow V}$.
In figure \ref{fig FU e BV} it is shown the time segment where the therapy is active. According to clinical results shown in \cite{Kabbi2005}, our model catches the relevant significance of addition of BV5 to the therapy: in this case TTP is increased from 7.5 months to 9.3 months  (in \cite{Kabbi2005} PFS increases from 7.3 to 9.2).
As explained in section \ref{section parameters}, some of the parameters whose values are missing, are estimated in a indirect way: we estimated these parameters, in order to obtain reasonable results of TTP (see fig 4 pag. in \cite{Kabbi2005}) and maintaining at the same time the biological constraints described in the previous section. In figure \ref{sens::TTP} we can appreciate how parameters influences effectiveness of the therapies. The rate of angiogenesis regression, as well as the reaction rate of angiogenesis to VEGF,  affects inversely the efficacy of therapy, in particular of bevacizumab. Moreover parameters linked to chemoresistant cells, $\eta_{res}, p$, act inversely on TTP, while the thickness of proliferating boundary acts proportionally, according to the fact that chemotherapy kills only proliferating cells. Although altering $(\lambda, \mu)$ will break the specific constraints imposed in subsection \ref{subsec::shape}, we include them into the analysis of sensitivity in order to show how growth rate influences TTP.

\begin{figure}[ptb]
\centering
\includegraphics[height=3.1168in,width=4.152in]
{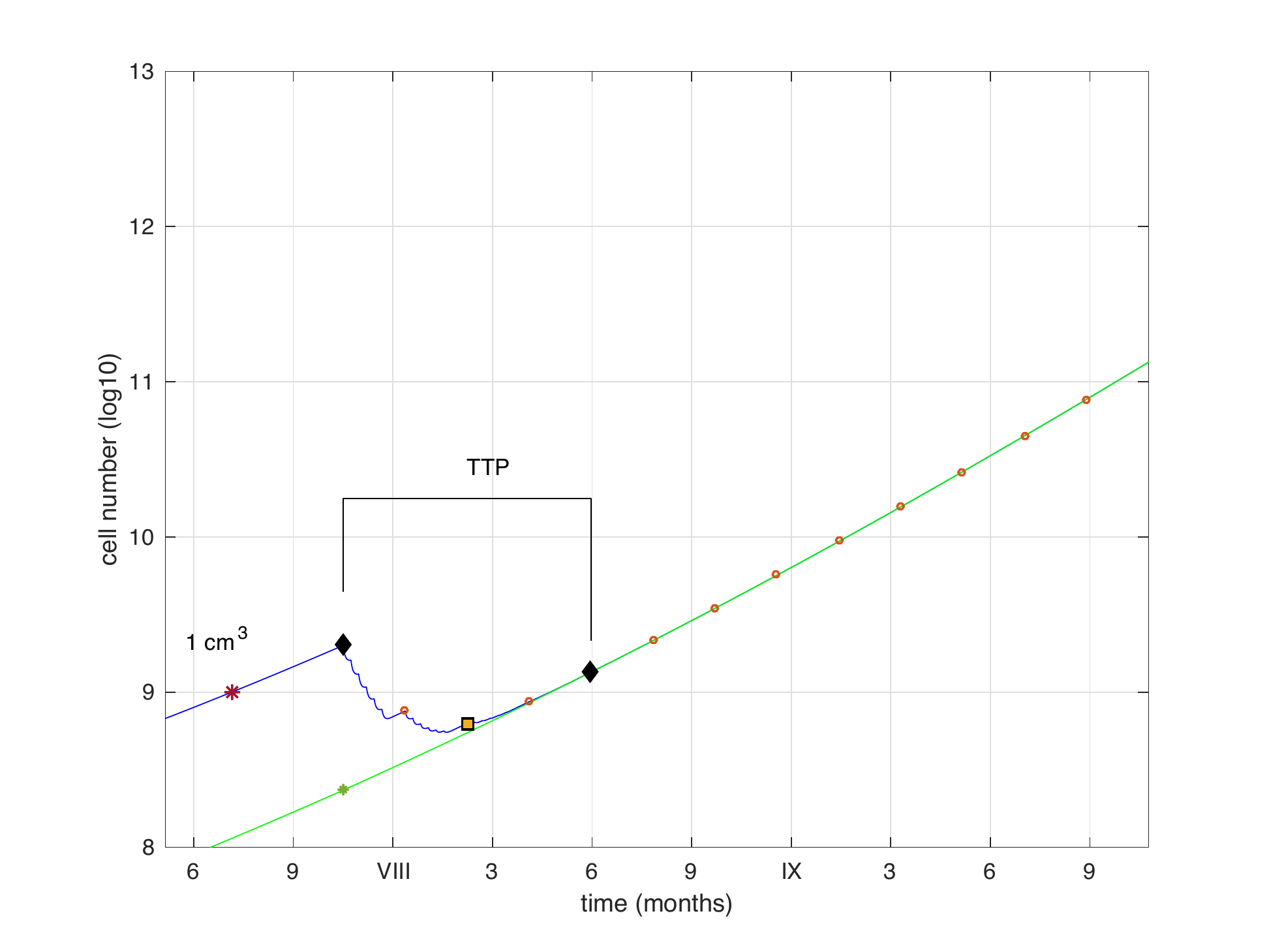}
\end{figure}

\begin{figure}[ptb]
\centering
\includegraphics[height=3.1168in,width=4.152in]
{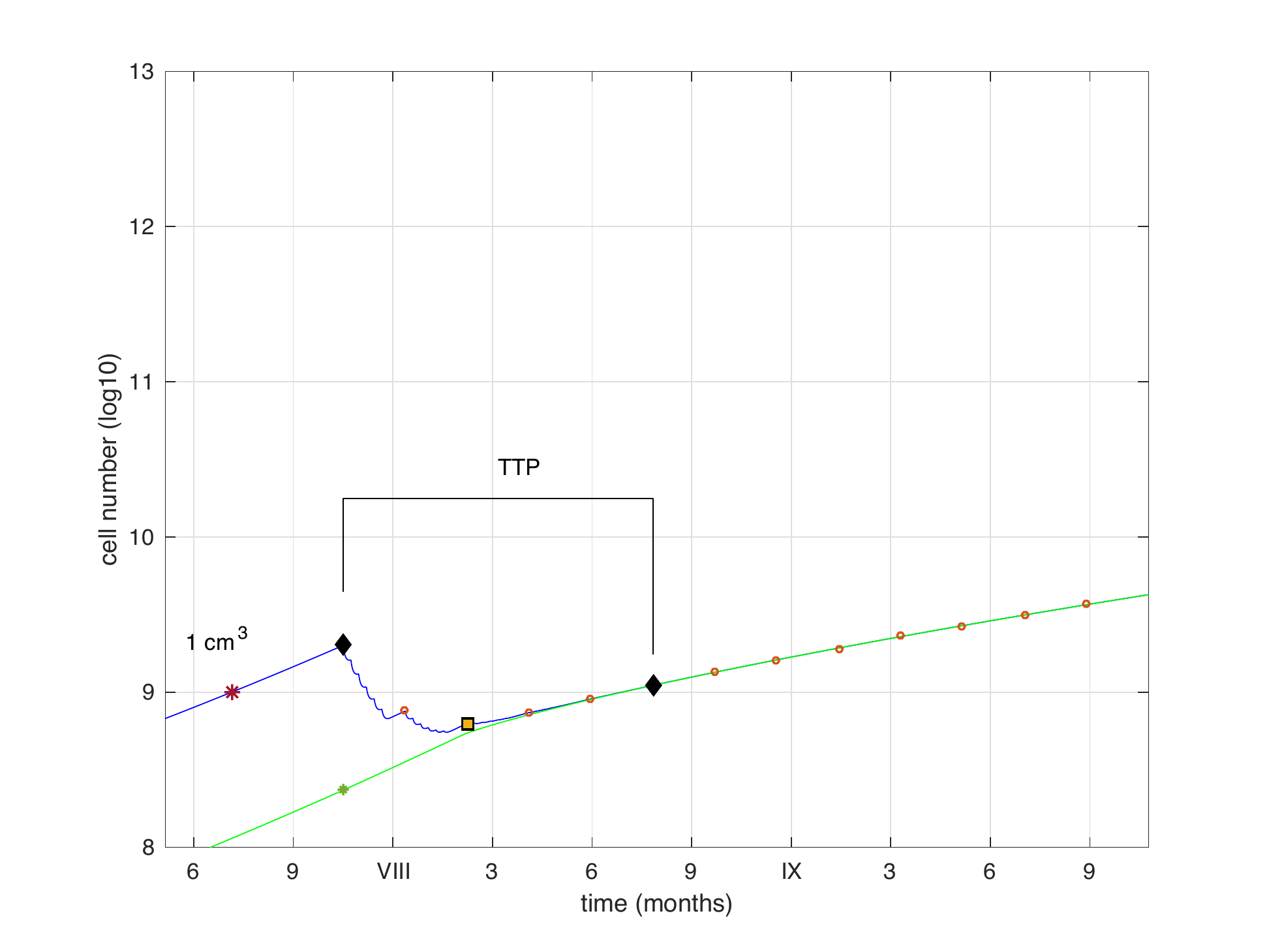}
\caption{{\protect\small Zoom around treatment section. The first figure show the treatment with only 5-FU while the second with 5-FU plus BV5. Orange circles denote time of assessment of tumor size, the bigger square represents the nadir point and the two black diamonds denote the limits for TTP.}}%
\label{fig FU e BV}
\end{figure}

\section{Conclusion}
We have devised a mathematical model that governs cancer growth, specifically in the case of solid tumors, and includes the action of a cytotoxic agent and of a monoclonal antibody as control terms. The equations considered here have been proved to be suitable to the case of colorectal cancer. Further investigation is required to claim their applicability to other kind of solid tumors, provided that the spherical approximation is reasonable. The action of the control functions could be used in a variety of contest where the effect of such treatment is required, but again, this will require additional effort to be established.

The model is formulated as a system of ordinary differential equations that describe the total number of cells, both sensible and resistant to the cytotoxic agent, the intensity of VEGF field and the level of vascularization due to angiogenesis. As expressed in the introduction the choice of VEGF as one of the variables was mainly due to our desire to include bevacizumab into the system. 
Even if the equations are ODEs we have taken into account the spatial structure through equations \eqref{eq::N^{sens}}, \eqref{eq::V_{t}}, \eqref{eq::N^{hypo}}: the rule used here for the number of proliferating cells  in absence of angiogenesis, namely $N_{boundary}^{norm}$, is a variant of Verhulst model, which has the form $N_{t}\left( 1-\left( \frac{N_{t}}{K} \right)^{\alpha}\right)$. We have devised this formula, based on geometrical considerations, to take into account the transition between an exponential growth and a slower growth when the number of cancerous cells is substantial. The 2/3 formula for VEGF was obtained from more physical argument, once more using the hypothesis of spherical symmetry. Due to the structure of the equations and their interaction we were able to obtain a curve of growth that results in a proper asymptotic rate in every phase of tumor development (figure \ref{fig::main}).

Numerical simulations were used both as a validation mechanism for the model and as a way to obtain an estimate for some of the parameters. In order to extract more information from the simulations, comparison with clinical data was also performed: we focused on the specific case of colorectal cancer and measure TTP following the treatment prescribed in \cite{Kabbi2005}. 
In this case we use TTP as an output to obtain additional informations on the parameters involved in the therapy.

The sets of parameters shown in table \ref{tab::parammathmodel} is able to match all the quantitative key landmarks imposed in section \ref{subsec::shape}, as well as TTP in both regimen 5FU and 5FU+BV5. 

In a subsequent work we shall investigate the modifications played by the explicit introduction of metastases, their growth, size at the time when therapy starts, impact on TTP computation. Another generalization we have
in mind, which moreover has been the initial motivation of our study, is to
investigate more complex therapies, as those summarized in
\cite{CremolFalcone}.  We have preferred to isolate the simplest case here because it may be modified to treat other types of cancer, due to its simplicity.


\begin{thebibliography}{99}                                                                                               %


\bibitem{Araklein} L. Arakelyan, Y. Merbl, P. Daugulis, Y. Ginosar, V. Vainstein, V. Selitser, Y. Kogan, H. Harpak, Z. Agur, Multi-Scale Analysis of Angiogenic Dynamics and Therapy, \textit{Cancer Modelling and Simulation}, 2003. 


\bibitem{Argyri}Katerina D. Argyri, Dimitra D. Dionysiou, Fay D. Misichroni, Georgios S. Stamatakos, Numerical simulation of vascular tumour
growth under antiangiogenic treatment:
addressing the paradigm of single-agent
bevacizumab therapy with the use of
experimental data, \textit{Biology Direct}, 2016, 11:12.

\bibitem{Agur} Agur, Z., Bloch, N., Gorelik, B., Kleiman, M., Kogan, Y., Sagi, Y.,  D. Sidransky, Ronen Y., Developing Oncology Drugs Using Virtual Patients of Vascular Tumor Diseases, \textit{Systems Bio in Drug Discovery}, pp. 201-237.




\bibitem {Benzekry}S. Benzekry, Mod\'{e}lisation et analyse math\'{e}matique
de th\'{e}rapies anti-canc\'{e}reuses pour les cancers m\'{e}tastatiques,
Ph.D. Thesis, Universit\'{e} de Provence, 2011.


\bibitem{BenzekryBenabdallah} S. Benzekry, D. Barbolosi, A. Benabdallah, F. Hubert, P. Hahnfeldt, Quantitative Analysis of the Tumor/Metastasis System and its Optimal Therapeutic Control, preprint. 

\bibitem{BenzekryHubert}S. Benzekry, G. Chapuisat, J. Ciccolini, A. Erlinger, F. Hubert,  A new mathematical model for optimizing the combination between antiangiogenic and cytotoxic drugs in oncology, \textit{C. R. Math. Acad. Sci. Paris} \textbf{350} (2012), n. 1-2, 23-28.

\bibitem {Bianco}A. R. Bianco, S. De Placido, G. Tortora, \textit{Core
curriculum. Oncologia clinica}, McGraw-Hill 2011.

\bibitem {Bolin}S. Bolin, E. Nilsson, R. Sj\"{o}dahl, Carcinoma of the colon
and rectum--growth rate, \textit{Annals of surgery} \textbf{198} (1983), n. 2, 151-158.

\bibitem {Bru}A. Br\`{u}, S. Albertos, J. L. Subiza, J. L. Garcia-Asenjo, I.
Br\`{u}, The universal dynamics of tumor growth, \textit{Biophys. J.}
\textbf{85} (2003), 2948-2961.

\bibitem {Chinnathambia}S. Chinnathambia, D. Velmuruganb, N. Hanagatad,
Investigations on the interactions of 5-fluorouracil with bovine serum
albumin: Optical spectroscopic and molecular modeling studies, \textit{Journal
of Luminescence} \textbf{151}, (2014), 1-10.

\bibitem {ColdmanMurray}A. J. Coldman, J.M. Murray, Optimal control for a
stochastic model of cancer chemotherapy, \textit{Math. Biosciences}
\textbf{168} (2000) 187-200.

\bibitem {CremolFalcone}C. Cremolini, M. Schirripa, C. Antoniotti, R. Moretto,
L. Salvatore, G. Masi, A. Falcone, F. Loupakis, First-line chemotherapy for
mCRC - a review and evidence-based algorithm, \textit{Nature Reviews Clinical
Oncology} \textbf{12} (2015), 607-619.

\bibitem {Donofrio}A D'Onofrio, A Gandolfi, A family of models of angiogenesis
and anti-angiogenesis anti-cancer therapy, \textit{Math. Medicine and Biology}
\textbf{26} (2009), n.1, 63-95.


\bibitem{DonofrioGandolfi} A D'Onofrio, A Gandolfi, Chemotherapy of vascularised tumours: role of vessel density and the effect of vascular "pruning", \textit{J Theor Biol.}, 2010 May 21;264(2):253-65.


\bibitem {Drake}J. W. Drake, B. Charlesworth, D. Charlesworth, J. F. Crow,
Rates of spontaneous mutation, \textit{Genetics} \textbf{148} (1998), 1667-1686.

\bibitem {Eisenhauer}E.A. Eisenhauer, P. Therasse, J. Bogaerts, L.H. Schwartz,
D. Sargent, R. Ford, J. Dancey, S. Arbuck, S. Gwyther, M. Mooney, L.
Rubinstein, L. Shankar, L. Dodd, R. Kaplan, D. Lacombe, J. Verweij, New
response evaluation criteria in solid tumours: Revised RECIST guideline,
European J. Cancer 45 (2009), 228-247.

\bibitem {Friberg}S. Friberg, S. Mattson, On the growth rates of human
malignant tumors: implications for medical decision making, \textit{Journal of
Surgical Oncology} \textbf{65} (1997), 284-297.

\bibitem{Gabhan} Mac Gabhann F1, Popel AS, Targeting neuropilin-1 to inhibit VEGF signaling in cancer: Comparison of therapeutic approaches, \textit{PLoS Comput Biol.}, 2006 Dec 29;2(12).



\bibitem {Gao}Z. Gao, M. J. Wyman, G. Sella, M. Przeworski, Interpreting the
dependence of mutation rates on age and time, PLOS Biology, 2016, 1-16.

\bibitem {Gaudreault}J. Gaudreault, G. Lieberman, F. Kabbinavar, V. Hsei,
Pharmacokinetics (PK) of bevacizumab (BV) in colorectal cancer,
\textit{Clinical Pharmacology and Therapeutics} 69 (2001).

\bibitem {Goldie}J. H.Goldie, A. J. Coldman, The genetic origin of drug
resistance in neoplasms: implications for systemic therapy, \textit{Cancer
Research} \textbf{44} (1984), 3643-3653.

\bibitem {Grady}W. Grady, J. M. Carethers, Genomic and epigenetic instability
in colorectal cancer pathogenesis, \textit{Gastroenterology} \textbf{135}
(2008), n. 4, 1079-1099-

\bibitem {Hahnfeld}P. Hahnfeldt, D. Panigraphy, J. Folkman, L. Hlatky, Tumor
development under angiogenic signaling: a dynamical theory of tumor growth,
treatment, response and postvascular dormancy, \textit{Cancer Research}
\textbf{59} (1999), 4770-4775.

\bibitem {HinowSwanson}P. Hinow, P. Gerlee, L. J. McCawley, V. Quaranta, M.
Ciobanu, S. Wang, J. M. Graham, B. P. Ayati, J. Claridge, K. R. Swanson, M.
Loveless, A. R. A. Anderson, A spatial model of tumor-host interaction:
application of chemotherapy, \textit{Math Biosci Eng.} \textbf{6} (2009), n.
3, 521-546.

\bibitem {Housman}G. Housman, S. Byler, S. Heerboth, K. Lapinska, M. Longacre,
N. Snyder, S. Sarkar, Drug Resistance in Cancer: An Overview, \textit{Cancers}
\textbf{6} (2014), 1769-1792.

\bibitem {Hurwitz}H. I. Hurwitz, L. Fehrenbacher, J. D. Hainsworth, W. Heim,
J. Berlin, E. Holmgren, J. Hambleton, W. F. Novotny, F. Kabbinavar,
Bevacizumab in combination with fluorouracil and leucovorin: an active regimen
for first-line metastatic colorectal cancer, \textit{J. Clin. Oncol.}
\textbf{23 }(2005), 3502--3508.


\bibitem{Jain} Jain HV1, Nör JE, Jackson TL., Quantification of endothelial cell-targeted anti-Bcl-2 therapy and its suppression of tumor growth and vascularization, \textit{Mol Cancer Ther.}, 2009 Oct;8(10):2926-36. 



\bibitem {Kabbi2003}F. F. Kabbinavar, H. I. Hurwitz, L. Fehrenbacher, N. J.
Meropol, W. F. Novotny, G. Lieberman, S. Griffing, E. Bergsland, Phase II,
randomized trial comparing bevacizumab plus fluorouracil (FU)/leucovorin (LV)
with FU/LV alone in patients with metastatic colorectal cancer, \textit{J.
Clin. Oncol.} \textbf{21} (2003), 60--65.

\bibitem {Kabbi2005}F. F. Kabbinavar, J. Schulz, M. McCleod, T. Patel, J. T.
Hamm, J. R. Hecht, R. Mass, B. Perrou, B. Nelson, W. F. Novotny, Addition of
Bevacizumab to Bolus Fluorouracil and Leucovorin in First-Line Metastatic
Colorectal Cancer: Results of a Randomized Phase II Trial, \textit{J. Clinical
Oncology} \textbf{23} (2005), n. 16, 3697-3705.

\bibitem {Kaldate}R. R. Kaldate, A. Haregewoin, C. E. Grier, S. A. Hamilton,
H. L. McLeod, Modeling the 5-Fluorouracil Area Under the Curve Versus Dose
Relationship to Develop a Pharmacokinetic Dosing Algorithm for Colorectal
Cancer Patients Receiving FOLFOX6, \textit{The Oncologist} 17 (2012), 296-302.

\bibitem {Komarova}N. Komarova, Stochastic modeling of drug resistance in
cancer, \textit{J. Theoret. Biology} \textbf{239} (2006) 351-366.


\bibitem{Klamka} Klamka J1, Maurer H, Swierniak A., Local controllability and optimal control for a model of combined anticancer therapy with control delays, \textit{Math Biosci Eng.}, 2017 Feb 1;14(1):195-216.

\bibitem{Landry} Landry, J., J. P. Freyer, and R. M. Sutherland. "A model for the growth of multicellular spheroids." Cell Proliferation 15.6 (1982): 585-594.

\bibitem{Ledzewicz} U. Ledzewicz, H. Schattler, Analysis of optimal controls for a mathematical model
of tumor anti-angiogenesis, \textit{Optim. Control Appl. Meth.}, 2006; 00:1–16


\bibitem {Machida}N. Machida, T. Yoshino, N. Boku, S. Hironaka, Y. Onozawa, A.
Fukutomi, K. Yamazaki, H. Yasui, K. Taku, M. Asaka, Impact of baseline sum of
longest diameter in target lesions by RECIST on survival of patients with
metastatic colorectal cancer, \textit{Japanese Journal of Clinical Oncology}
\textbf{38} (2008), n.10, 689-694.

\bibitem {Marme}D. Marm\'{e}, N. Fusenig Editors, \textit{Tumor Angiogenesis},
Springer-Verlag Berlin 2008.

\bibitem {Motl}S. Motl, Bevacizumab in Combination Chemotherapy for Colorectal
and Other Cancers, \textit{American J. Health-System Pharmacy} \textbf{62
}(2005), n.10, 1021-1032.


\bibitem{Simeoni}Monica Simeoni, Paolo Magni, Cristiano Cammia, Giuseppe De Nicolao, Valter Croci, Enrico Pesenti, Massimiliano Germani, Italo Poggesi, Maurizio Rocchetti, Predictive pharmacokinetic-pharmacodynamic modeling of tumor growth kinetics in xenograft models after administration of anticancer agents.
\textit{Cancer Res.} 2004 Feb 1; 64(3): 1094–1101.


\bibitem{Stefanini} Stefanini MO, Qutub AA, Mac Gabhann F, Popel AS., Computational models of VEGF-associated angiogenic processes in cancer, \textit{Math Med Biol.}, 2012 Mar;29(1):85-94.


\bibitem {Perthame}B. Perthame, Some mathematical models of tumor growth, 2015.

\bibitem {Peters}G. J Peters, J. Lankeimal, R. M. Kok, P. Noordhuis, C. J. van
Groeningen, C. L. van der Wilt, S. Meyer, H. M Pinedo, Prolonged retention of
high concentrations of 5-fluorouracil in human and murine tumors as compared
with plasma, \textit{Cancer Chemother Pharmacol} \textbf{31} (1993), 269-276.

\bibitem {Riganti}C. Riganti, E. Mini, S. Nobili, Editorial: Multidrug
resistance in cancer: pharmacological strategies from basic research to
clinical issues, \textit{Front. Oncol.} 2015.

\bibitem {Sadahiro}S. Sadahiro, T. Suzuki, K. Ishikawa, T. Nakamura, Y.
Tanaka, K. Ishizu, S. Yasuda, H. Makuuchi, C. Murayama, Estimation of the time
of pulmonary metastasis in colorectal cancer patients with isolated
synchronous liver metastasis, \textit{Japan J. Clin. Oncol.} \textbf{35}
(2005), n.1, 18-22.

\bibitem {Simeoni}M. Simeoni, P. Magni, C. Cammia, G. De Nicolao, V. Croci, E.
Pesenti, M. Germani, I. Poggesi, and M. Rocchetti. Predictive
pharmacokinetic-pharmacodynamic modeling of tumor growth kinetics in xenograft
models after administration of anticancer agents. Cancer Res., 64 :1094--1101,
Feb 2004.

\bibitem {Steel}G. G. Steel, \textit{Growth kinetics of tumor}, Oxford:
Clarendon Press, 1977.

\bibitem {TalkingtonDurrett}A. Talkington, R. Durrett, Estimating tumor growth
rates in vivo, \textit{Bull. Math. Biology} \textbf{77} (2015), n. 10, 1934-1954.

\bibitem {Therasse}P. Therasse, S. G. Arbuck, E. A. Eisenhauer, J. Wanders, R.
S. Kaplan, L. Rubinstein, J. Verweij, M. Van Glabbeke, A. T. van Oosterom, M.
C. Christian, S. G. Gwyther, New guidelines to evaluate the response to
treatment in solid tumors, \textit{J. National Cancer Institute} \textbf{92}
(2000), n.3, 205-216.

\bibitem {Tlsty}T.D. Tlsty, B.H. Margolin, K. Lum, Differences in the rates of
gene amplification in nontumorigenic and tumorigenic cell lines as measured by
Luria--Delbruck fluctuation analysis, \textit{Proc. Natl Acad. Sci. USA}
\textbf{86} (1989), n. 23, 9441-9445.

\bibitem {Umetani}N. Umetani, T. Masaki, T. Watanabe, S. Sasaki, K. Matsuda,
T. Muto, Retrospective radiographic analysis of nonpedunculated colorectal
carcinomas with special reference to tumor doubling time and morphological
change, \textit{The American Journal of Gastroenterology} \textbf{95} (2000), 1794--1799;

\bibitem {Weinberg}R. A. Weinberg, \textit{The Biology of Cancer}, second
edition, Garland Science, Taylor and Francis Group, New York 2014.

\bibitem {Zhao}B. Zhao, S. M. Lee, H.-J. Lee, Y. Tan, J. Qi, T. Persigehl, D.
P. Mozley, L. H. Schwartz, Variability in Assessing Treatment Response:
Metastatic Colorectal Cancer as a Paradigm, \textit{Clin Cancer Res};
\textbf{20} (2014), n. 13, 3560-3568.



\end{thebibliography}
\end{document}